\documentclass[a4paper]{jpconf}
\usepackage{graphicx}
\usepackage{cite}
\usepackage{amssymb,amsmath}
\usepackage{psfrag}
\usepackage[usenames]{color}
\begin{document}
\title{Lateral distributions of electrons in air showers initiated by ultra-high energy gamma quanta taking into account LPM and geomagnetic field effects}

\author{Tatyana Serebryakova$^1$, Alexander Goncharov$^1$, Anatoly Lagutin$^1$, Roman Raikin$^1$ and Akeo Misaki$^2$}

\address{$^{1}$ Altai State University, 61 Lenin Ave., Barnaul 656049, Russia}
\address{$^{2}$ Saitama University, 255 Shimookubo, Saitama-city 338-8570, Japan}

\ead{serebryakova@theory.asu.ru}

\begin{abstract}
Lateral distributions of electrons in air showers initiated by photons of ultra high energies ($10^{18}-10^{22}$~eV) obtained on the basis of numerical solution of adjoint cascade equations are presented. An extended analysis has been made considering separately the Landau-Pomeranchuk-Migdal (LPM) effect and the interaction of photons and electrons with the geomagnetic field (GMF) with respect to the scaling formalism for lateral distributions. It is shown that one-parametric scaling description of the lateral distribution of electrons remains valid up to the highest energies considering the LPM and GMF effects, that allow effective primary particle type discrimination using the surface detectors data of largest ground-based air shower arrays.
\end{abstract}

\section{Introduction}
The determination of the ultra-high energy (UHE) cosmic ray composition, including identification of primary photons and setting up high-precision limits on their fluxes is a vital effort for understanding the origin of cosmic rays~\cite{Aloisio1,Aloisio2,Risse1,HomolaPAO:2018,Karpikov:2018}. UHE photons could originate either in the interactions of primary cosmic-ray nuclei with the cosmic microwave background radiation  or in the exotic phenomena, e.g. resulting from decay of hypothetical super-massive particles. Though the latter scenarios are highly disfavoured by the results of recent studies, the reliable identification of primary photons from the experimental data of giant air shower arrays on both the average and the event-by-event basis remains an actual problem.

The limits on UHE photons fluxes established so far are very stringent, well below the predictions of the most of the exotic models. Thus, extracting the information about the primary particle type from the data of surface detectors, which provides an order of magnitude larger statistics in comparison with atmospheric telescopes, plays an important role.

While simulations of nuclei-induced air showers at ultra-high energies rely on hadronic interaction models fundamentally uncertain in the relevant energy range, in case of primary photons the following two effects should be properly taken into account: the Landau-Pomeranchuk-Migdal (LPM) effect~\cite{Landau:1953/1, Landau:1953/2, Migdal:1956} and the interaction of EHE photons and electrons with the geomagnetic field (GMF)~\cite{ McBreen-Lambert}. In this paper we present the results of the extended analysis of lateral distributions of electrons in photon-generated air showers of $10^{18}-10^{22}$ eV considering separately the LPM and GMF effects with respect to the scaling formalism for lateral distributions. We show that one-parametric scaling description of the lateral distribution of electrons remains valid up to the highest energies considering the LPM and GMF effects. Effectiveness of primary particle type discrimination using the scale factors of lateral distributions measured by surface detectors of largest ground-based air shower arrays is discussed.

\section{Calculation methods and results of analysis}
The numerical solution of adjoint cascade equations was used to calculate characteristics of electromagnetic cascades in the atmosphere and magnetosphere of the Earth~\cite{Gonch:1991, PlyaAharonian:2002,Gonch:2003,Gonch:2005}.

We consider all the essential processes of cascade particle interaction with matter at energy region $E\geq 10^4$ eV. Calculations were performed for the US standard atmosphere. The GMF intensity profile corresponds to the southern location of the Auger Observatory (Mendoza, Argentina). The primary energies range is $E_{\gamma}=10^{18}-10^{22}$ eV.

\begin{figure}[t]
%\psfrag{Eg, eV}[l][l][2]{$E_{\gamma}$, eV}
%\psfrag{Nge(E,t)}[l][l][2]{$N_{\gamma,e}(E,t)$}
%\begin{center}
\hspace{-1.0cm}
\begin{minipage}{10cm}
\includegraphics[angle=-90, width=1\textwidth]{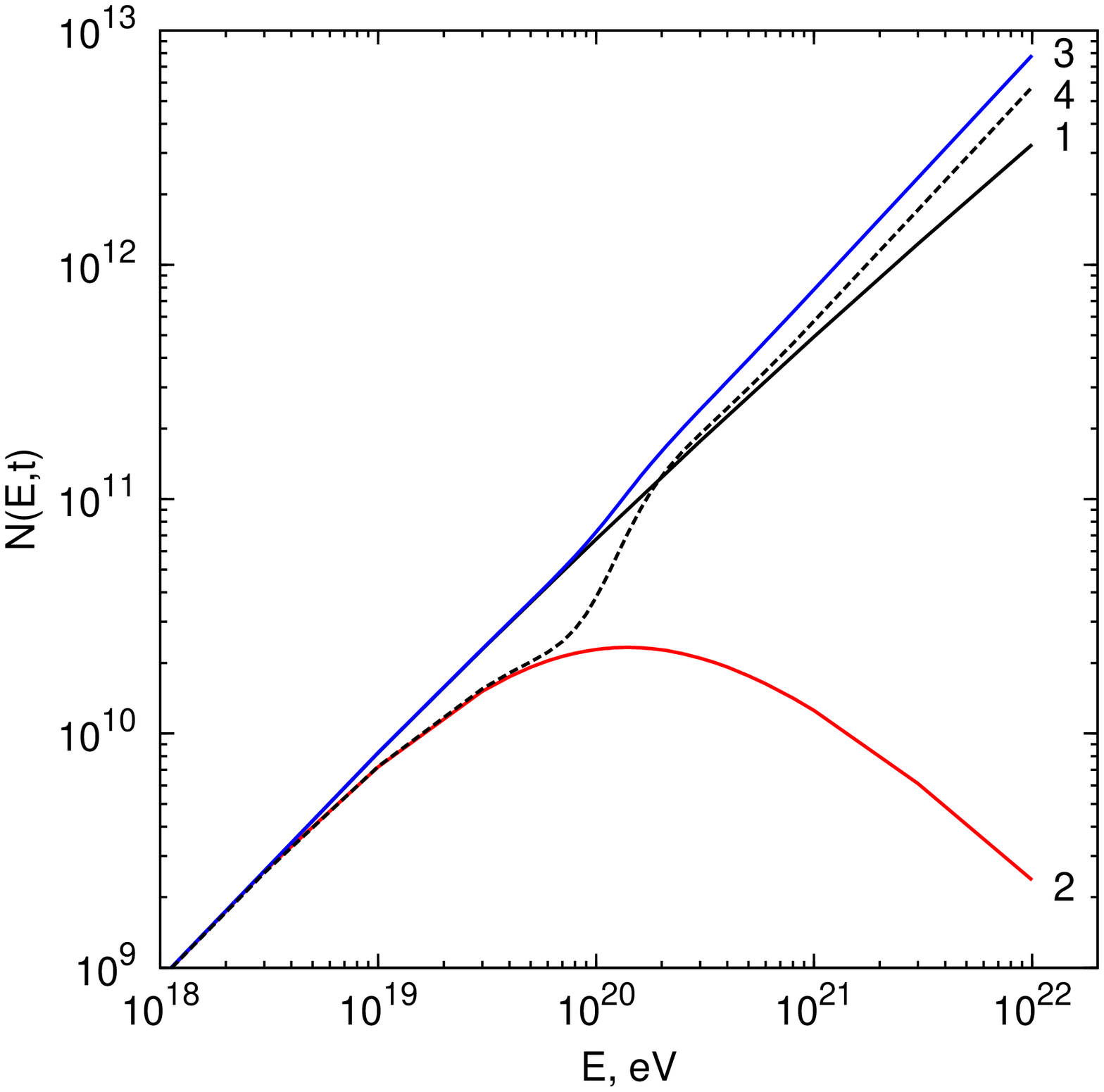}
%\\ \hspace{1cm} a)
\end{minipage}
\hspace{-2cm}
\begin{minipage}{10cm}
\includegraphics[angle=0, width=1\textwidth]{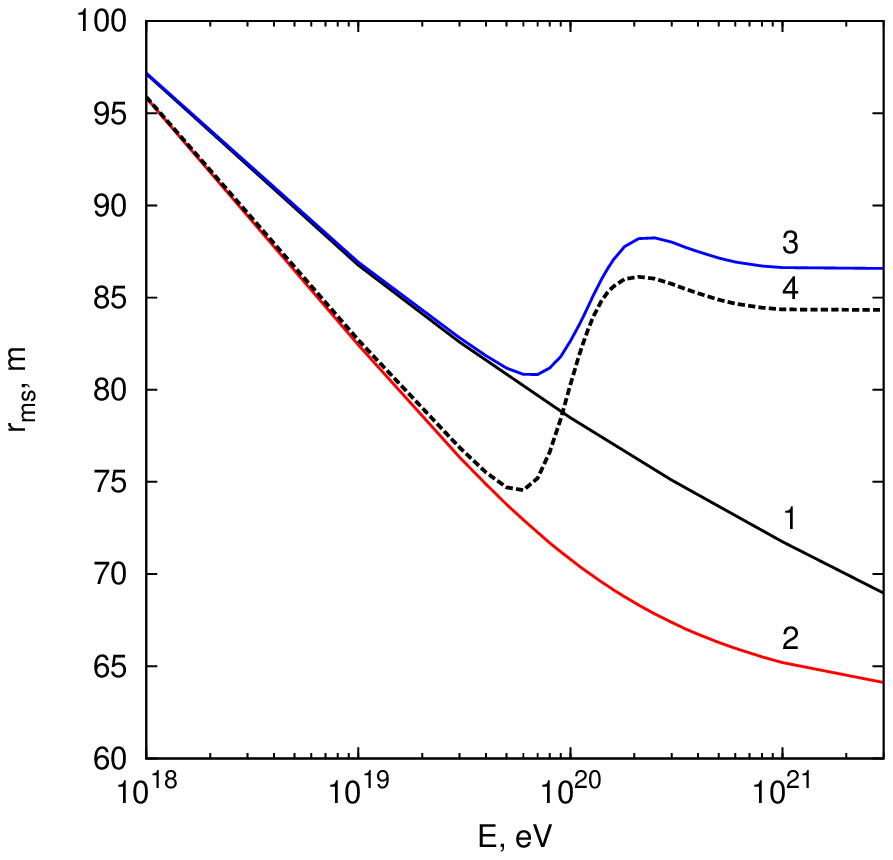}
%\\ \hspace{2cm} b)
\end{minipage}
%\hspace{.1cm}
\begin{minipage}{20cm}
\hspace{.5cm} a) \hspace{7.5cm} b)\\
(1) \full ~without effects \quad (3) \textcolor{blue}{\full} only GMF\\
(2) \textcolor{red}{\full} only LPM \quad\quad\quad (4) \dashed ~LPM+GMF\\
\end{minipage}
\vspace{-.5cm}
\caption{\label{ne_total}Energy dependences of the total number of electrons (a) and root mean square radius of electron component (b) in vertical electromagnetic air showers produced by photons with $E_{\gamma}=10^{18}-10^{22}$ eV at the observational level of $890$~g/cm$^2$ with and without taking into account the UHE-effects (LPM and GMF).} 
%The total number of electrons on the observation level $890$ g/cm$^2$ for vertical \mbox{$\gamma$-produced} cascade.}
%\end{minipage}
%\end{center}
\end{figure}

In Figure~\ref{ne_total} energy dependences of the total number of electrons $N(E,t)$ (shower size) and root mean square radius of electron component in vertical electromagnetic air showers produced by photons with $E=10^{18}-10^{22}$ eV at the observational level of $t=890$~g/cm$^2$ are shown with and without taking into account the UHE-effects (LPM and GMF).

It can be seen from Figure~\ref{ne_total} that the interaction with GMF counteracts the LPM effect by converting the primary photon into a bunch of synchrotron photons with energies effectively below the threshold of the LPM effect~(see~\cite{PlyaAharonian:2002} for details).

The root mean square radius of electron component $r_{\rm ms}$ is defined in a standard way as a second moment of electron lateral distribution function (LDF):
\begin{equation}
r_{\rm{ms}}(E,t)=\left(\displaystyle\frac{2\pi}{N(E,t)}\int_0^\infty r^2 \rho(r;E,t)rdr\right)^{1/2},
\end{equation}
where $\rho(r;E,t)$ is the local particle density at radial distance $r$ from the core position in shower with primary energy $E$ observed at the depth $t$ in the atmosphere.

According to the scaling approach~\cite{Lagutin:1997}, when radial distance is rescaled with respect to the variable $x=r/r_{\rm ms}$, the shape of electron LDF doesn't depend on primary energy and shower age for $x=0.05-20$:
\begin{equation}
\rho(r;E,t)=\frac{N(E,t)}{r^2_{\rm ms}}F\left(\frac{r}{r_{\rm ms}}\right).
\label{scal2}
\end{equation}
Here $F(x)$ is the invariant function. Thus, $r_{\rm ms}$ could be considered as a single parameter describing variations of the shape of electron LDF with energy and observation depth. It is important that, according to our calculations, the scaling property~(\ref{scal2}) remains valid up to the highest energies regardless of what UHE-effects were taken into account. Numerical values of the ratios of $r_{\rm ms}^{\ast}$ with taken into account UHE-effects to $r_{\rm ms}$ are presented in Table~\ref{ratio_rms}.

Later it was shown~\cite{Lagutin:1997D,Lagutin:2001,LagutinL2002} that the scaling property of the same type extends to electron LDF in extensive air showers generated by primary nuclei. 

In order to examine the possibility for effective primary particle discrimination with identifications of photons on average and event-by-event basis we have simulated vertical air showers generated by photons, protons and iron nuclei of $10^{18}$~eV and evaluated root mean square radii from lateral distributions of electrons and also depths of maximum $t_{\rm max}$. Simulations were performed using  CORSIKA v.7.4100~\cite{CORSIKA:1998} with EPOS LHC v.3400 (FLUKA 2011.2c.2) hadronic interaction models. In order to get reliable data on local particle densities at very large distances from the shower core the thinning level and particle weight limit were set as $\varepsilon_{\rm th}= 10^{-8}$ and $\omega=10^2$ respectively.

\begin{figure}[h]
a)\includegraphics[width=21pc]{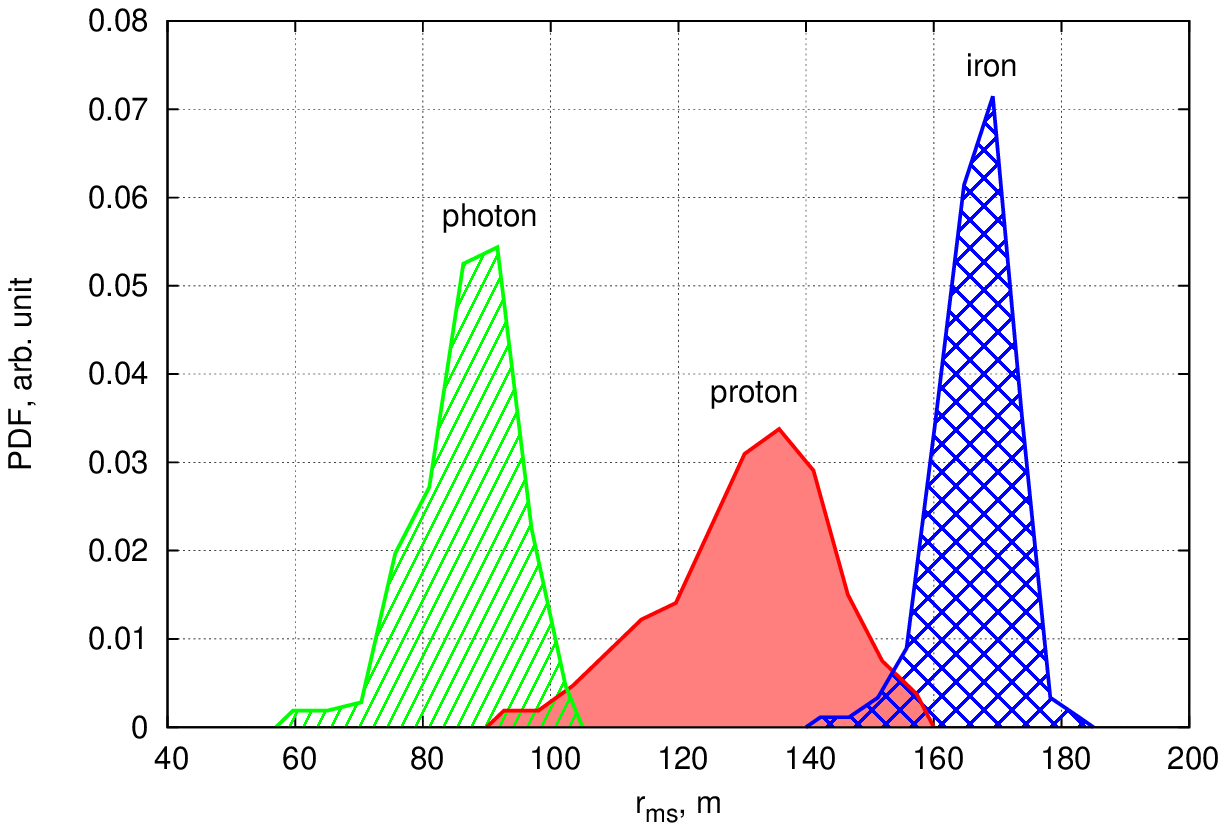}\\ \hspace{2pc}%
b)\includegraphics[width=21pc]{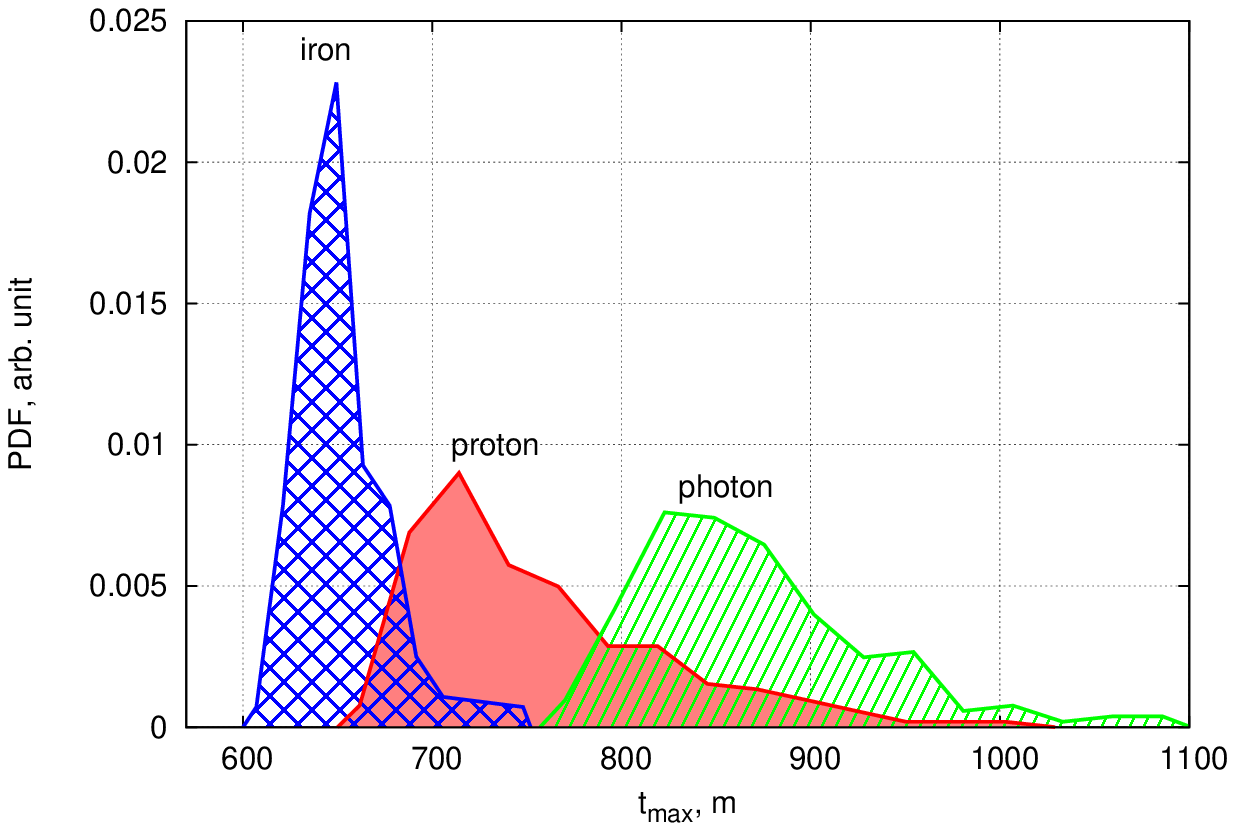}\hspace{2pc}%
\begin{minipage}[b]{14pc}\caption{\label{distr}
Distributions of primary particle discriminating observables: root mean square radius of electron component $r_{\rm ms}$ at sea level (a) and depth of shower maximum $t_{\rm max}$ (b) for vertical air showers initiated by photons (green), protons (red) and iron nuclei (blue) with primary energy $10^{18}$~eV.}
\end{minipage}
\end{figure}

The distributions of $r_{\rm ms}$ and $t_{\rm max}$ for different primaries are displayed in Figure~\ref{distr}. 200 showers were included in each data set. One can see that the root mean square radius of electron component is generally more sensitive to the primary particle type in comparison with the depth of shower maximum.

\begin{table}[h]
\caption{\label{ratio_rms} The behavior of the ratio of $r_{\rm ms}^{\ast}$ with taken into account effects to $r_{\rm ms}$ without effects for vertical cascade on the bservation level $890$ g/cm$^2$.}
\begin{center}
\begin{tabular}{cccc}
\br
$E_{\gamma}, eV$&$\frac{r_{\rm ms}^{LPM}}{r_{\rm ms}}$&$\frac{r_{\rm ms}^{GMF}}{r_{\rm ms}}$&$\frac{r_{\rm ms}^{LPM+GMF}}{r_{\rm ms}}$\\
\mr
$10^{18}$       & 0.987 &1.000& 0.987\\
$10^{19}$       & 0.950 &1.002& 0.953\\
$3\cdot 10^{19}$& 0.924 &1.003& 0.931\\
$10^{20}$       & 0.902 &1.053& 1.023\\
$3\cdot10^{20}$ & 0.897 &1.172& 1.142\\
$10^{21}$       & 0.909 &1.207& 1.176\\
$3\cdot10^{21}$ & 0.930 &1.255& 1.223\\
$10^{22}$       & 0.958 &1.309& 1.274\\
\br
\end{tabular}
\end{center}
\end{table}

\section{Conclusion}
Using the numerical solution of adjoint cascade equations we analysed characteristics of air showers produced by ultra-high energy photons taking into account the Landau-Pomeranchuk-Migdal effect and the interaction of UHE photons and electrons with the geomagnetic field.
An extended analysis has been made considering separately the LPM and GMF effects with respect to the scaling formalism for lateral distributions of electrons. It was obtained that one-parametric scaling description of the lateral distribution of electrons remains valid up to the highest energies. CORSIKA Monte-Carlo simulations of air showers initiated by primary photons and nuclei showed that using root mean square radius of electron component under the framework of scaling approach allows effective primary particle type discrimination from surface detectors data of largest ground-based air shower arrays.

\ack 
This work was supported by Russian Foundation for Basic Research (grant 16-02-01103 a).

\end{document}